%% file: main.tex
\documentclass[acmsmall]{acmart}

\usepackage{enumitem}
\usepackage{xcolor}

\newcommand\rev[1]{{\textcolor{black}{#1}}} 

\newcommand\todo[1]{{\rev{[TODO: {#1}]}}}

\newcommand\revv[1]{{\textcolor{black}{#1}}} 

\newcommand\revi[1]{{\textcolor{black}{#1}}} %
\newcommand\revii[1]{{\textcolor{black}{#1}}} %
\newcommand\reva[1]{{\textcolor{black}{#1}}} %

\newboolean{hidecomments} 
\setboolean{hidecomments}{false}

\ifthenelse{\boolean{hidecomments}}
{\newcommand{\cb}[2]{}}
{\newcommand{\cb}[2]{
    \fbox{\bfseries\sffamily\scriptsize#1}
    {\sf\small$\blacktriangleright$ %
      {#2} $\blacktriangleleft$}}} %

\ifthenelse{\boolean{hidecomments}}
{\newcommand{\ctd}[2]{}}
{\newcommand{\ctd}[2]{\todo[color=cyan,inline]{#1: #2}}}

\AtBeginDocument{%
  \providecommand\BibTeX{{%
    \normalfont B\kern-0.5em{\scshape i\kern-0.25em b}\kern-0.8em\TeX}}}

\setcopyright{acmlicensed}
\acmJournal{PACMHCI}
\acmYear{2025} \acmVolume{9} \acmNumber{2} \acmArticle{CSCW071} \acmMonth{4}\acmDOI{10.1145/3710969}

\begin{document}

\title[Real-time Feedback of Non-verbal Cues in Meetings]{Exploring a Real-time Feedback Display of Non-verbal Cues in Online Work Meetings to Support Self-Presentation}

\author{Kevin Chow}
\authornote{Both authors contributed equally to this research.}
\email{kchowk@cs.ubc.ca}
\orcid{0009-0008-5455-3328}
\affiliation{%
  \institution{Department of Computer Science, University of British Columbia}
  \city{Vancouver}
  \state{BC}
  \country{Canada}
}

\author{Roy Rutishauser}
\authornotemark[1]
\email{rutis@ifi.uzh.ch}
\orcid{0000-0002-1416-9830}
\affiliation{%
  \institution{Department of Informatics, University of Z{\"u}rich}
  \city{Z{\"u}rich}
  \country{Switzerland}
}

\author{Andr\'e~N.~Meyer}
\email{ameyer@ifi.uzh.ch}
\orcid{0000-0002-5532-1181}
\affiliation{%
  \institution{Department of Informatics, University of Z{\"u}rich}
  \city{Z{\"u}rich}
  \country{Switzerland}
}

\author{Joanna McGrenere}
\email{joanna@cs.ubc.ca}
\orcid{0000-0002-8411-3293}
\affiliation{%
  \institution{Department of Computer Science, University of British Columbia}
  \city{Vancouver}
  \state{BC}
  \country{Canada}
}

\author{Thomas Fritz}
\email{fritz@ifi.uzh.ch}
\orcid{0000-0002-1834-6240}
\affiliation{%
  \institution{Department of Informatics, University of Z{\"u}rich}
  \city{Z{\"u}rich}
  \country{Switzerland}
}

\renewcommand{\shortauthors}{Kevin Chow, Roy Rutishauser, André N. Meyer, Joanna McGrenere \& Thomas Fritz}

\begin{abstract}
\input{sections/abstract}
\end{abstract}



\maketitle

\input{sections/1-introduction}

\input{sections/2-relatedwork}

\input{sections/3-novecs}

\input{sections/4-method}

\input{sections/5-findings}
\input{sections/6-discussion}
\input{sections/7-conclusion}
\begin{acks}
The authors thank the anonymous reviewers, members of the UZH HASEL and the UBC eDAPT Lab for their valuable and insightful feedback. In particular, we would like to thank Remy Egloff for their involvement in developing the initial research prototype from which \textit{Novecs} took inspiration. This work was supported by funding from \textit{Designing interactive digital technologies for a post-pandemic world} (RGPIN-2023-045457) and \textit{SNSF TaskFlow Grant} (207916). 
\end{acks}

\received{January 2024}
\received[revised]{July 2024}
\received[accepted]{October 2024}

\bibliographystyle{ACM-Reference-Format}
\bibliography{ref}

\end{document}

%% file: sections/abstract.tex
Expressing oneself appropriately in online meetings through non-verbal cues can be challenging for knowledge workers. 
Automatic non-verbal cue detection technologies have the potential to support workers’ self-presentation efforts through real-time feedback, but little is known about workers' reactions to and the implications of doing so. 
We designed and implemented \textit{Novecs} as a technology probe of a real-time feedback display that automatically detects and signals users' own non-verbal cues --- smiling, nodding, gaze, and posture. \textit{Novecs} was deployed in an exploratory field study (n=18) to support knowledge workers' self-presentation in their everyday meetings. 
Post-study interviews reveal how \textit{Novecs'} real-time feedback helped increase in-the-moment self-awareness, and how neutrally-framed feedback may \revii{help navigate} tensions between authentic and in-authentic self-presentation.
\revii{Participants also emphasized the need for natural timing when adjusting non-verbal cues in-meeting.}
We discuss design opportunities and challenges of real-time, non-verbal cue feedback systems, such as personalizing feedback based on different meeting types.

%% file: sections/1-introduction.tex
\section{Introduction}

\revv{Meetings are a regular occurrence for most knowledge workers, but they come with a host of challenges. Some of these challenges are exacerbated in online meetings, which have become a cornerstone of remote and hybrid work.}
One such challenge is in non-verbal self-presentation \cite{goffman2002presentation}, namely how workers convey information about themselves to others through non-verbal cues such as facial expressions or body language in order to shape how they are perceived. 
Non-verbal cues are important for signaling engagement, communicating more effectively, building trust and rapport with colleagues, and resolving conflicts \cite{standaert2021shall, gemmell2000gaze, neureiter2013hand, kostic2014social, remland2016nonverbal, vinciarelliSocialsignalstheir2008, pereiraCommunicationSkillsTraining2021}.

Online meetings have unique considerations --- limited video field-of-view \cite{walther2015interpersonal, bailenson2021nonverbal}, default-on self-view \cite{kuhn2022constant}, latency \cite{hinds1999cognitive}, and multitasking \cite{marlow2016taking} --- that can require increased self-awareness and more intentional user effort, ultimately contributing to ``non-verbal overload'' \cite{bailenson2021nonverbal} and ``Zoom fatigue'' \cite{fauville2023video, fauville2021zoom, walther2007selective, bergmann2023meeting}. 
\revii{
Online meeting attendees often need to exaggerate their nodding to ensure their signal of agreement is perceived \cite{bailenson2021nonverbal,choi2021kairos}.
\reva{The work of knowledge workers typically involves the discussion and exchange of knowledge and ideas through meetings. As a result, they may face specific challenges in meetings, where their focus may be divided between often cognitively demanding work-related discussions and their self-presentation efforts \cite{wicklundInfluenceSelfAwarenessHuman1979}.}
Consciously monitoring their own self-view to maintain appropriate non-verbal cues in such meetings can be difficult; focusing too much on one's own self-view may also distract from their work or even contribute to Zoom fatigue \cite{kuhn2022constant}. 
While self-viewing can enhance self-awareness \cite{miller2017through,k2021meeting}, on its own, it may not be able to help those who lack knowledge of but desire to exhibit appropriate non-verbal cue behaviour \cite{aliLISSALiveInteractive2015,aliSocialskillstraining2017}.
Additionally, self-viewing by itself cannot intervene or notify users of their non-verbal cues in real-time, requiring constant monitoring for self-presentation. 
}

\revv{
There is an opportunity to leverage technological approaches to address the aforementioned challenges of self-awareness and cognitive load when it comes to non-verbal self-presentation in online meetings.
Automatic non-verbal cue detection technology continues to mature, achieving robust, real-time processing speeds even on more generalized hardware like laptop webcams \cite{baltrusaitis2018openface,bianco2016robust,martinez2016advances}. 
Interactive feedback systems that support the automatic detection and display of real-time feedback on one's own non-verbal cues could help facilitate self-awareness and \revi{self-reflection} on self-presentation during online work meetings. 
}

\revv{
While such real-time, non-verbal cue feedback systems may have the potential to support self-presentation in unique and powerful ways, there are open questions that should first be explored before expending effort into building robust systems \cite{klasnjaHowevaluatetechnologies2011}.
Perspectives around self-presentation may be especially nuanced for knowledge workers in online work meetings, necessitating careful consideration when it comes to the design of the feedback display. 
Might users feel like they are being judged?
Furthermore, the importance and appropriateness of non-verbal cue behaviour may depend on factors related to the type of meeting a knowledge worker is attending (e.g., one-on-one versus group meetings or who the other meeting attendees are) as well as individual differences regarding self-presentation. Real-time feedback systems, which are often designed to be glanceable, have known challenges around distractibility \cite{kulyk2006real, leshed2009visualizing, tanveer2015rhema}, but work meetings may present context-specific considerations and challenges. Privacy is another key concern --- some knowledge workers already face intrusive, possibly privacy-impeding, monitoring at work \revi{\cite{bergerProductivityTheatre, das2023algorithmic,constantinides2022good}}, especially those working remotely --- how might privacy be preserved in such systems?}

Exploring these questions can further inform and inspire \revii{the} design of future non-verbal cue feedback systems, and improve our understanding of the potential of such technologies for supporting knowledge workers. 
To do so, we adopted the use of a \textit{technology probe}, which is a technological artifact that ``[collects] information about the use and the users of the technology in a real-world setting'' \cite{hutchinson2003technology}. We felt that an in-situ approach, where knowledge workers could experience the technology in the cadence of their everyday online work meetings, was critical for surfacing the nuanced perspectives, opportunities, and challenges of real-time, non-verbal cue feedback systems.  

\rev{
\revv{Thus, w}e designed and implemented \textit{Novecs} (short for \textbf{No}n-\textbf{ve}rbal \textbf{c}ue \textbf{s}ystem) as a technology probe \cite{hutchinson2003technology} that \revv{enables the automatic detection and real-time feedback display of users' own non-verbal cues during their online work meetings.} 
\textit{Novecs} targets four types of kinesic \cite{hartzler2014real, liuImprovingMedicalStudents2016} non-verbal cues, including two of facial expressions --- smiling and nodding --- and two of body language --- posture and gaze.
\revv{While we are primarily interested in the potential of the real-time, in-meeting feedback, \textit{Novecs} secondarily also includes summative, post-meeting feedback that visualizes users' non-verbal cues after each meeting.}
\revv{\textit{Novecs} accounts for privacy concerns by processing all data \textit{locally} on users' machines.}
}
In addition, we designed the feedback to be \revv{neutral and avoid being normative} so that users would not feel judged by the probe \rev{and would be encouraged to reflect on their non-verbal behaviour in the moment.} 
\revv{Altogether,} these design decisions reflect our intention for \textit{Novecs} to empower and support the user, rather than be used against them. 

Our work aimed to answer the following research questions:
\begin{itemize}
    \item \textbf{RQ1:} \revv{How do knowledge workers leverage real-time, non-verbal cue feedback in their various online work meetings?}
    \item \textbf{RQ2:} \revv{How might the use of a real-time, non-verbal cue feedback system reveal and/or influence knowledge workers' perspectives on their non-verbal self-presentation in online work meetings?} 
    \item \textbf{RQ3:} \revv{What are the opportunities and challenges of a real-time, non-verbal cue feedback system for online work meetings to support non-verbal self-presentation?}
\end{itemize}

To answer these research questions, we deployed \textit{Novecs} in a multi-week exploratory field study with 18 knowledge workers, who used it in their real-world online work meetings. In the first, pre-feedback phase of the study, \textit{Novecs} was deployed without signaling any feedback (only showing user presence/absence) for at least five meetings to allow participants to acclimate themselves to having it running with their regular work meetings. Subsequently, during the feedback phase, participants experienced \textit{Novecs} in at least ten online work meetings each, while receiving in- and post-meeting feedback.

\revv{
Findings from exit interviews show that \textit{Novecs} helped increase participants' awareness of their non-verbal cues and supported their desire to express their non-verbal cues with appropriate and natural timing. \textit{Novecs'} neutral and non-normative feedback design may have supported the surfaced tensions between authentic and in-authentic self-presentation. The perceived utility of \textit{Novecs'} real-time feedback also varied based \revi{on} different meeting types.
We discuss the importance of authenticity and how participants valued being imperfect to be more ``human'', as well as balancing noticeability and distractibility through a consideration of meeting types.}

\revv{Overall,} this paper makes the following contributions:
\begin{itemize}
  \item \textbf{Empirical} \textbf{findings} from deploying the \textit{Novecs} technology probe in-situ with knowledge workers' online work meetings that highlight the nuanced opportunities and challenges of a real-time feedback display of non-verbal cues, including tensions in \revv{supporting both authentic and in-authentic self-presentation}. 
  \item \rev{\textbf{Design} \textbf{opportunities} for \rev{neutral and non-normative} real-time feedback of non-verbal cues, such as an increased consideration of meeting types, \revv{utilizing personas for customizing non-verbal cue feedback}, and accepting imperfection in non-verbal cue behaviour.} 
\end{itemize}

%% file: sections/2-relatedwork.tex
\section{Background and Related Work}

\revv{
In this section, we first provide a brief overview of non-verbal self presentation and the real-time, automatic detection of non-verbal cues. We then delve into related work on interactive non-verbal cue feedback systems for contexts outside of meetings, before concluding with systems designed specifically for online work meetings. 
}

\subsection{Self-Presentation and Non-verbal Cues}

\revii{
Self-presentation refers to the ways that one attempts to control how they are perceived by others \cite{goffman2002presentation}. 
}
\revii{While individuals engage in self-presentation efforts in everyday social interactions with those around them, the unique implications of the work context make it particularly interesting -- such as in the formation of one's professional identity and its role in organizational and team dynamics \cite{bolino2016impression}.}
\revi{Self-presentation at work} has been widely studied, including its impact on leadership, trust, and worker well-being \revi{\cite{ellemers2006social,arkin1989self,baumeister1989motives,baumeister1987self,peck2018acting,klotz2018good,dubrin2010impression,gardner1988impression,yuksel2017brains}}. Self-presentation efforts can be both controlled or automatic as well as authentic or inauthentic \cite{bolino2016impression,schlenker2003self}. 
Authentic self-presentation is when one's outer expressions are congruent with their inner state, whereas in-authentic self-presentation is when they are incongruent \cite{goffman2002presentation,peck2018acting}. An example of controlled, in-authentic self-presentation in a work meeting might be when one consciously attempts to look engaged and alert (such as through non-verbal cues like frequent nodding and sitting up straight) even when feeling tired. 
Non-verbal self-presentation \cite{depaulo1992nonverbal} specifically involves non-verbal cues, such as facial expressions and body language. %
\rev{In this work, we focus on self-presentation in online work meetings as a large proportion of social interactions between work colleagues now occurs on videoconferencing platforms such as Zoom, Teams, or Google Meet.}
Online meetings can be especially challenging for self-presentation, as ``users are forced to consciously monitor non-verbal behaviour and to send cues to others that are intentionally generated'' \cite{bailenson2021nonverbal}, often leading to increased cognitive load and ``Zoom fatigue'' \cite{fauville2023video, fauville2021zoom, walther2007selective, bergmann2023meeting, shockley2021fatiguing}.

\subsection{Automatic Detection of Non-verbal Cues in Real-time}

\revi{Researchers have made advancements in enabling automatic, real-time detection of non-verbal cues, using both computer vision techniques as well as other sensing modalities (e.g., specialized sensors \cite{kim2008meeting, terken2010multimodal}, mobile phones and wearables, such as smartwatches or earables \cite{choi2021kairos}). In particular, computer vision algorithms are now able to detect non-verbal cues in real-time, while using standard camera hardware like laptop webcams.}
\revv{Studies have explored the detection of the four non-verbal cues we chose to explore through \textit{Novecs} in real-time} --- posture \cite{dmelloMultimodalsemiautomatedaffect2010,dmelloPosturePredictorLearner}, gaze \cite{dmelloGazetutorgazereactive2012}, smiling \cite{bianco2016robust,howeDesignDigitalWorkplace2022}, and nodding \cite{artiran2021hmm, tan2003real} --- and also others like head shakes \cite{kapoor2001real} and eye blinks \cite{soukupovaRealTimeEyeBlink}.  
\revii{
Such ``lower-level'' non-verbal cues are specific, observable behaviours. They have been used in aggregate to help predict ``higher-level constructs'', which broadly try to encapsulate the complexities of an individual's state of being (e.g., emotions). 
\textit{Novecs'} four ``lower-level'' non-verbal cues are often associated with desirable ``higher-level'' constructs in the meeting context, such as engagement (posture, gaze), agreement (nodding), and positivity (smiling).   
Examples of aggregation to predict ``higher-level constructs'' include research in emotion recognition \cite{dmelloMultimodalsemiautomatedaffect2010, novielliSensorBasedEmotionRecognition}, for detecting engagement when watching television, advertisements, and in the classroom \cite{hernandezMeasuringengagementlevel2013,jaquesPredictingAffectIntelligent, mcduffPredictingAdLiking2015, babaeiFacesFocusStudy2020}, detecting fatigue \cite{jabbarDriverDrowsinessDetection2020}, and for mediating group discussions \cite{byun2011honest, kim2008meeting}. 
}

\revii{In our work, \textit{Novecs} maintains the use of computer-vision algorithms with standard laptop webcams (hence not requiring any additional hardware) to detect ``lower-level'' non-verbal cues rather than ``higher-level'' constructs.
Predicting ``higher-level'' constructs can face additional challenges with accuracy and objectivity, especially across diverse use cases and cultural demographics \cite{noauthor_microsoft_nodate}. 
Notably, feedback on ``higher-level'' constructs may be more difficult to act upon \cite{cook2020interventions}.
\reva{Some may struggle with grasping how they should change their behaviour to improve a ``high-level'' construct like engagement whereas specific ``lower-level'' non-verbal cues like posture are more easily acted upon, especially in real-time during an already cognitively demanding meeting \cite{faucett2017should}. ``Lower-level'' cues can also face challenges of accuracy and cognitive load in meetings, and we explore their potential for in-meeting feedback through \textit{Novecs}.}}

\subsection{Displaying Feedback on Non-verbal Cues Outside of Online Work Meetings}

\revv{Existing work has utilized automatic detection technologies to enable interactive feedback systems for improving users' non-verbal self-presentation in a variety of contexts such as media or job interviews \cite{pereiraCommunicationSkillsTraining2021,hemamouSlicesAttentionAsynchronous2019,hoqueMACHmyautomated2013,rasipuramcomprehensiveevaluationaudiovisual2019}, telehealth \cite{liuImprovingMedicalStudents2016,faucett2017should}, or just for general conversational skills in non-work contexts \cite{aliLISSALiveInteractive2015,hoqueRichNonverbalSensing2014,rasipuramAutomaticmultimodalassessment2020,aliSocialskillstraining2017}.
These existing systems resulted in some improvements in participants' self-presentation, suggesting the potential of such systems. However, the majority were evaluated in controlled lab settings and only provided \textit{summative feedback} on users' non-verbal cues. Real-time feedback may have unique opportunities and challenges akin to those of glanceable displays, such as enabling in-the-moment awareness and adjustments but also challenges with distraction \cite{kulyk2006real,leshed2009visualizing, tanveer2015rhema,tausczik2013improving}. 
}

\revv{
Researchers in telehealth are among the few to explore \textit{real-time feedback} of non-verbal cues to support clinician self-presentation in patient-clinician video calls, which can be considered a specific instance of an online work meeting. Hartzler et al. \cite{hartzlerRealtimeFeedbackNonverbal2014} conducted a design study with a Wizard-of-Oz real-time feedback system, demonstrating that an ambient, glanceable approach is acceptable to clinicians for displaying real-time information about their non-verbal cues. 
Faucett et al. \cite{faucett2017should} implemented ReflectLive, a functional system for patient-clinician video calls that includes glanceable, real-time, non-verbal cue feedback on clinicians' speaking contribution, eye gaze, conversational overlap, and screen centredness. Through a user evaluation with 10 clinicians, they found that ReflectLive increased self-awareness of eye gaze and was not overly distracting \cite{faucett2017should}.
}

While \textit{Novecs} also adopts a real-time feedback approach, it was designed for a more general context of online work meetings rather than patient-clinician video calls. There are key differences --- in the telehealth context, patient-clinician video calls are fairly structured and well-defined: they are one-on-one, the clinician plays a specific role in the conversation and is the one that receives non-verbal cue feedback, and desirable non-verbal cue behaviour is well-studied (i.e., observational studies have linked non-verbal cues with desirable patient/clinical outcomes \cite{faucett2017should, mast2007importance}). On the other hand, a knowledge worker might attend a wide range of different types of work meetings even within a single work day, including one-on-ones or larger group meetings and with varying other meeting attendees. 
\revi{This variety in knowledge workers' meeting types suggests that desirable non-verbal self-presentation may differ from meeting to meeting, and is far more nuanced than in patient-clinician video calls, where a more normative approach might be appropriate.
Knowing this, we deviated from existing work by deliberately taking a more neutral and non-normative approach to presenting feedback, in order to better accommodate the diverse range of possible meetings that workers might engage in.}

\subsection{Interactive Feedback Systems for Online Work Meetings}

\revv{In the context of online work meetings, prior work has explored a variety of interactive systems for supporting knowledge workers, providing feedback on various factors, including both verbal and non-verbal cues, such as speaker participation \cite{samroseCoCoCollaborationCoach2018, kim2008meeting}, attendee experience \cite{asenieroMeetCuesSupportingOnline2020}, inclusion
 and relational affiliation~\cite{cutlerMeetingEffectivenessInclusiveness2021a, hartzler2014real}, speaker consensus \cite{samroseMeetingCoachIntelligentDashboard2021}, and emotions \cite{garcia1999, ezzaouia2020emodash, samroseCoCoCollaborationCoach2018}.}
Particularly relevant is Samrose et al.'s \cite{samroseMeetingCoachIntelligentDashboard2021} MeetingCoach, as it includes a subset of the same non-verbal cues as \textit{Novecs}. The tool provides an automated post-meeting dashboard that summarizes relevant information, including speaking turn and consensus, but also non-verbal cues like head nods and facial sentiment. They found that the MeetingCoach dashboard helped improve meeting attendees' awareness of meeting dynamics, and that they desired more ``feedback assistance to make meetings more effective and inclusive'' \cite{samroseMeetingCoachIntelligentDashboard2021}.  
However, MeetingCoach, like much of the existing work on interactive feedback systems for non-verbal cues, only provides feedback in a summative fashion, leaving unknown the potential value of real-time feedback for promoting in-the-moment self-awareness of non-verbal cues. 
\revi{Real-time feedback allows users to not only be aware of and adjust their non-verbal cues while a meeting is taking place, but also supports reflection on their self-presentation \textit{in-the-moment} of the meeting in a more granular manner.}

\revv{
To the best of our knowledge, we are one of the first to explore automatic detection and display of real-time, non-verbal cue feedback in a neutral, non-normative manner with knowledge workers' real (thus highly varied) online work meetings to both support and better understand self-presentation in this context.  
}

%% file: sections/3-novecs.tex
\section{The \textit{Novecs} Technology Probe} 

\textit{Novecs} (short for \textbf{No}n-\textbf{ve}rbal \textbf{c}ue \textbf{s}ystem) is a technology probe \cite{hutchinson2003technology} designed to explore automatic, real-time feedback for supporting non-verbal self-presentation in online work meetings.
\revi{\textit{Novecs} was designed to encourage an individual knowledge worker to personally reflect on and become more aware of their non-verbal cues during the varied range of online meetings they attend for work. Rather than enforcing specific behaviour and to be more inclusive, \textit{Novecs} avoids overly normative design in non-verbal cue feedback and considers the individual's goals for their self-presentation. Our design goals evolved in conjunction with various rounds of piloting that we had conducted with graduate students and other knowledge workers who used multiple iterations of \textit{Novecs} with their own meetings, with some using it up to a week. We also elicited feedback on the design of \textit{Novecs} from an hour-long session with expert HCI researchers from one of the lead authors' institution. Piloting helped us to refine our final set of non-verbal cues, adjust interface and cue feedback design, and make improvements to the detection algorithm for each of the cues (see Section~\ref{sec:detectingnvcues}).}

\revi{In the following sections, we first describe our rationale behind selecting the four non-verbal cues, how these cues are detected, how meeting sessions are managed, and end with how real-time feedback and summative feedback are displayed through the In-Session Sidebar and the Post-Session Summary window, respectively.}

\revi{\subsection{Selection of Non-verbal Cues}}

\revi{
We considered several factors in our selection of non-verbal cues for the \textit{Novecs} technology probe, including visibility in a video feed as well as accuracy and technical feasibility without needing specialized hardware (i.e., can be run locally on laptops and laptop webcams). 
Overall, \textit{Novecs} supports the sensing of four kinesic \cite{hartzler2014real,liuImprovingMedicalStudents2016} non-verbal cues, also known as movements of the head and body, including both facial expressions and body language: posture, gaze, smiling, and nodding. 
We focused on kinesic behaviours, as they are among the most salient in the teleconferencing context \cite{liuImprovingMedicalStudents2016} -- frequently expressed, changing within the field-of-view of the video feed, can be easily noticed by other meeting attendees, and also be easily changed by the user themselves.
}
\revii{
In addition, these four cues also all have a range of well-documented and accurate sensing algorithms in literature \cite{dmelloMultimodalsemiautomatedaffect2010,dmelloPosturePredictorLearner,dmelloGazetutorgazereactive2012,howeDesignDigitalWorkplace2022,artiran2021hmm,tan2003real}, that are feasible with regular laptop webcams, crucial for deployment in the field in a study like ours. 
}

\revi{
We decided not to include other potentially relevant kinesic cues like eye blinks, eyebrow raises, or hand gestures, \revii{as they are either not very} visually salient, have challenges with sensing accuracy \cite{liuImprovingMedicalStudents2016,hartzler2014real}, or can be difficult for an individual to be aware of or change.
Existing work has also looked at other relevant non-verbal cues like speaking time or tone, but analyzing those would require additional access to the audio stream of the online work meeting, and thereby potentially capturing meeting content. We chose not to do so to avoid intruding on participants' (and other meeting attendees') privacy and to facilitate recruitment.}

\revi{\subsection{Detection of Selected Non-Verbal Cues}\label{sec:detectingnvcues}}

\textit{Novecs} analyzes image frames with pre-trained machine learning models using OpenFace 2.0 \cite{baltrusaitis2018openface} to detect non-verbal cues. We use pre-trained models to allow for the image processing to run locally on users’ machines, and delete image frames after processing. No specialized hardware is required, as the processing supports any regular built-in or external webcam. %
OpenFace \cite{baltrusaitis2018openface} has built-in facial landmark detection that enables the calculation of presence and intensity information for action units in the Facial Action Coding System (FACS) \cite{ekman1978facial}. 
\revi{It} also outputs head position, rotation, and eye gaze information, all of which we use for detecting non-verbal cues.
\reva{Occlusion and low illumination conditions present accuracy challenges for any vision-based system, and may be exacerbated in real-world scenarios. Despite these common limitations, OpenFace has been shown to be robust even in non-ideal lighting, when faces are non-frontal or partially occluded. OpenFace has also been evaluated on multiple datasets comprising samples of subjects with diverse ethnic backgrounds and skin colors (e.g., \cite{zhang15_cvpr, klare2015pushing}).}
\textit{Novecs} is implemented as an Electron app with Node.js v16.17.x and supports macOS 12+. 
Detected non-verbal cue signals are stored in an SQLite database residing on users' local machine.

To later visualize the detected non-verbal cues to the user in the In-Session Sidebar (section~\ref{sec:inmeetingsidebar}) and Post-Session Summary (section~\ref{sec:postmeetingsummary}), the sensed values need to be compared against a user's target threshold, their individualized ideal behavior for that non-verbal cue.
Users personalize their target thresholds through two calibration steps, a per-individual and per-session calibration.
The \textit{per-individual} calibration is performed once with the initial setup of \textit{Novecs} to capture per-user thresholds. Users can set these thresholds themselves to personalize the sensitivity of each feature (e.g., allowing for more or less mobility on their chairs).
The \textit{per-session} calibration is run at the beginning of each meeting to determine ``target gaze and posture'' for the particular meeting, and can also help to account for changes in different physical meeting environments.

The following list describes the sensing approach of the four non-verbal cues:

\begin{itemize}
	\item \textbf{Posture:} 3D head position is used as a proxy for users' posture as we found from piloting that this works well as a low-cost, accurate estimation of posture. The distance between the target head position (from the per-session calibration) to the detected head position is evaluated against a user-defined threshold. \textit{Novecs} detects non-target posture when the distance exceeds the threshold. The direction of non-target posture (backwards, forwards, sideways) is based on the direction in which the target and detected head positions differ. 

	\item \textbf{Gaze:} Gaze is calculated by considering both eye gaze and head rotation in a similar manner as posture. Incorporating head rotation was found in piloting to produce a more robust detection of gaze, as users often turn their heads when looking in a different direction. Both the difference between target and detected eye gaze and head rotation are calculated and each compared against user-defined thresholds to produce an eye gaze and head rotation score. This score is then combined and compared against a system-defined threshold to decide if a non-target gaze signal (directionless) should be detected. 

	\item \textbf{Smiling:} A smile is detected and displayed if the sum of the intensity of AU6 (cheek raiser) and intensity of AU12 (lip corner puller) \cite{friesen1983emfacs, howeDesignDigitalWorkplace2022} is greater than a threshold for 500 milliseconds or longer. The threshold was defined based on piloting as the minimum time to detect a smile without too many false positives, e.g. when a person is speaking.

	\item \textbf{Nodding:} To detect nodding, our approach examines the pitch (vertical head rotation across the x-axis) of a set of frames in a time window of 1 second (typical duration of a nod according to \cite{Kawato2000}) to determine if a nod has occurred. Specifically, it checks this set of frames for a ``high-low-high'' pattern in the frame sequence, first by finding the lowest the head is rotated and subsequently checking for the highest points before and after. If both differences (from the lowest head rotation) exceed the user-defined nod angle threshold, and also a movement score (standard deviations of translation components of the head position) remains below a user-defined movement threshold, a ``nod'' signal is displayed. We found that accounting for the general movement of the head helped reduce false positives.

\end{itemize}

\begin{figure}[h]
  \centering
  \includegraphics[width=1.0\textwidth]{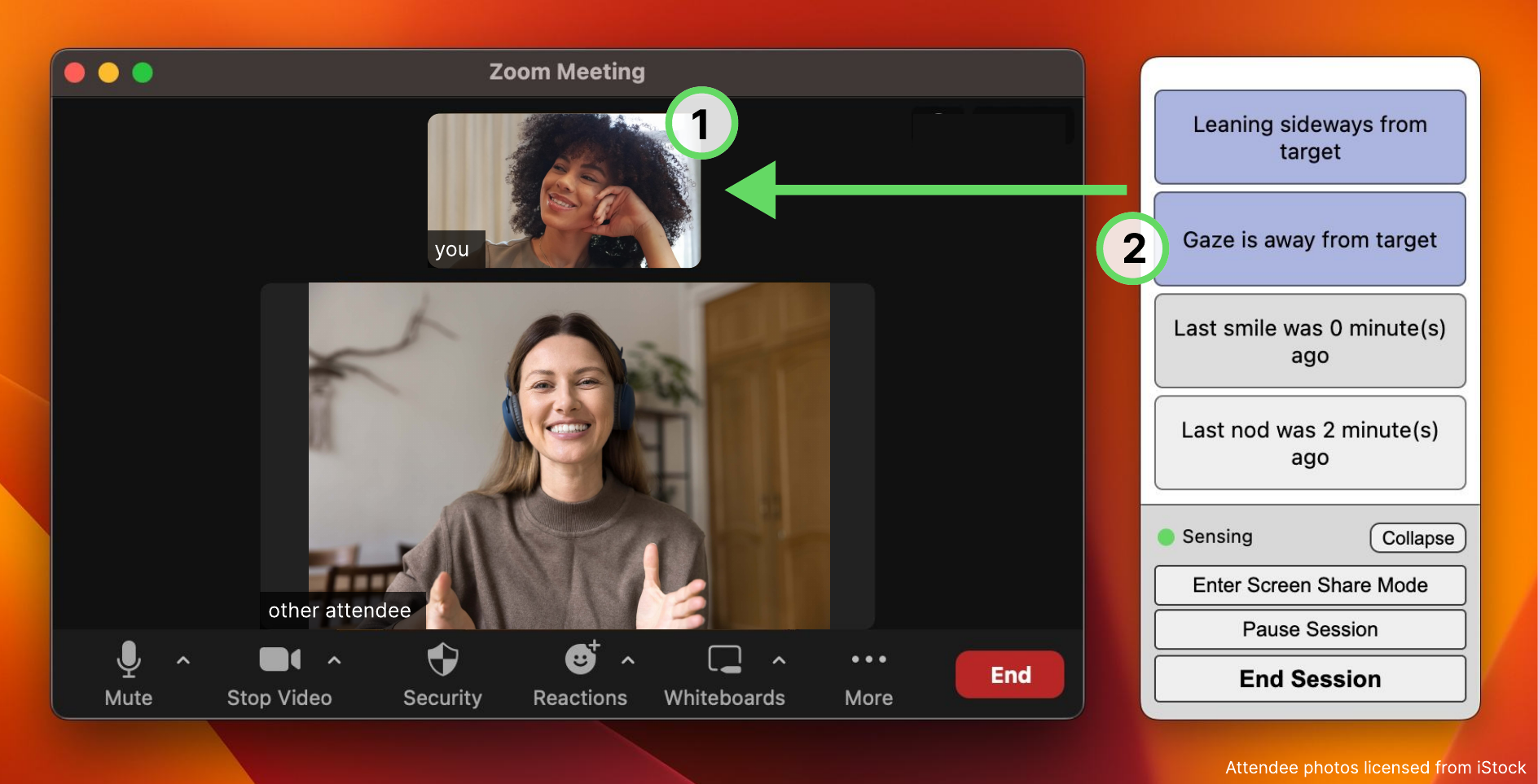}
  \caption{\reva{Screenshot of the \textit{Novecs} In-Session Sidebar (2) used in parallel to a Zoom video meeting. It indicates that the user ((1) denoted as `you') is currently displaying non-target posture (leaning sideways, widget changed to blue), non-target gaze (closed eyes, widget changed to blue), is currently smiling (widget changed to grey, last smile counter reset to 0 minutes), and recently nodded (2 minutes ago).}}
  \label{fig:mt-sidebar}
\end{figure}

\subsection{Managing \textit{Novecs} Sessions}
For each meeting, users can decide if they want to manually start \textit{Novecs}, either through the menubar or through a small reminder notification that briefly appears when Zoom or Microsoft Teams are active (i.e., window is in focus). When triggered, a pop-up window appears prompting users to estimate the length of the meeting (used for an end-of-meeting reminder notification) and choose the webcam for the meeting. In this paper, we use ``session'' to refer to when the \textit{Novecs} probe is enabled for a meeting.
\rev{As \textit{Novecs} does not ask for admin access (to avoid further restricting potential participants), it does not have access to whether or not a webcam is being used and by which applications, meaning that \textit{Novecs} cannot automatically start or stop a session based on the status of an online meeting and whether or not the user's webcam is on or off.}

\subsection{In-Session Sidebar}\label{sec:inmeetingsidebar}

Once a \textit{Novecs} session is started, the In-Session Sidebar appears and initiates the \textit{per-session} calibration step after a short 3-second countdown. \rev{This calibration step, which prompts users to maintain their ``target posture and gaze'', lasts until ten calibration frames are captured (the duration of which will depend on the performance of the user’s machine; calibration times typically range from 2-5 seconds).} In the study, we explained to participants that they themselves are free to define what their own ``target'' posture and gaze was (i.e., their own ideal or preference). 

The In-Session Sidebar (see Figure~\ref{fig:mt-sidebar}) consists of four vertically stacked feedback widgets, one for each of the four non-verbal cues, along with controls for collapsing the sidebar (collapsing only makes it smaller to \rev{save} space), entering/exiting screen sharing mode for users to toggle when sharing their screen (turns sidebar into miniature version to avoid sharing the feedback with other meeting attendees), pausing the session (pauses non-verbal cue detection), and ending the session. The In-Session Sidebar is also implemented to remain always-on-top while a session is started to display glanceable feedback. 
\revi{While an ideal implementation would have the widgets embedded into one's own self-view on video-conferencing platforms, we did not pursue this option for technical reasons and to allow participants to use their own \revii{choice} of video-conferencing platforms for work. Nonetheless, we encouraged study participants to position the sidebar close to their video-conferencing platform window where their self-view would be located, like in Figure~\ref{fig:mt-sidebar}.}
A status indicator displays if the non-verbal cue detection is currently active, paused, or does not detect a user in frame.

The feedback widgets provide glanceable, real-time feedback to users on each of the four non-verbal cues via changes in colour and text. We outline how the widgets display feedback for each non-verbal cue below:

\begin{itemize}
	\item \textbf{Posture:} Colour changes from grey (target posture) to blue and text changes to indicate the direction in which \textit{Novecs} detected out-of-bound posture (backwards, forwards, sideways). 
	\item \textbf{Gaze:} Colour changes from grey (target gaze) to blue and text changes to signify that the user has been looking away from the \rev{target gaze area} for a significant amount of time. 
	\item \textbf{Smiling/Nodding:} Colour switches from grey to a darker grey for the duration of the sensed smile/nod. The text is updated to display the number of minutes since the last sensed smile/nod. 
\end{itemize}

\rev{Core to the design of the feedback widgets was our design goal of encouraging nuanced self-reflection by avoiding an overly normative framing. 
To ameliorate concerns about feeling judged from piloting early iterations of \textit{Novecs}, we decided to use more neutrally-associated colours, blue/grey instead of red/green, and less normative language, using a self-defined, per-user ``target'' instead of language such as ``(not) ideal'', which may imply one ideal standard of non-verbal cue expression.
\revii{We additionally ruled out red/green to be more inclusive towards colourblind individuals.}
Displaying the number of minutes since last smile/nod (rather than directly prompting users to smile/nod) was another deliberate design decision to avoid artificially pushing users to smile/nod, while still facilitating self-awareness of existing smiling/nodding behaviour. 
We chose to use a less salient colour change for smiling and nodding (dark grey) compared to posture and gaze (blue) as users may benefit from being alerted of non-target posture/gaze in real-time during their meeting, whereas they likely do not need to be alerted of when they smiled or nodded.}

\begin{figure}[t]
  \centering
  \includegraphics[width=0.90\textwidth]{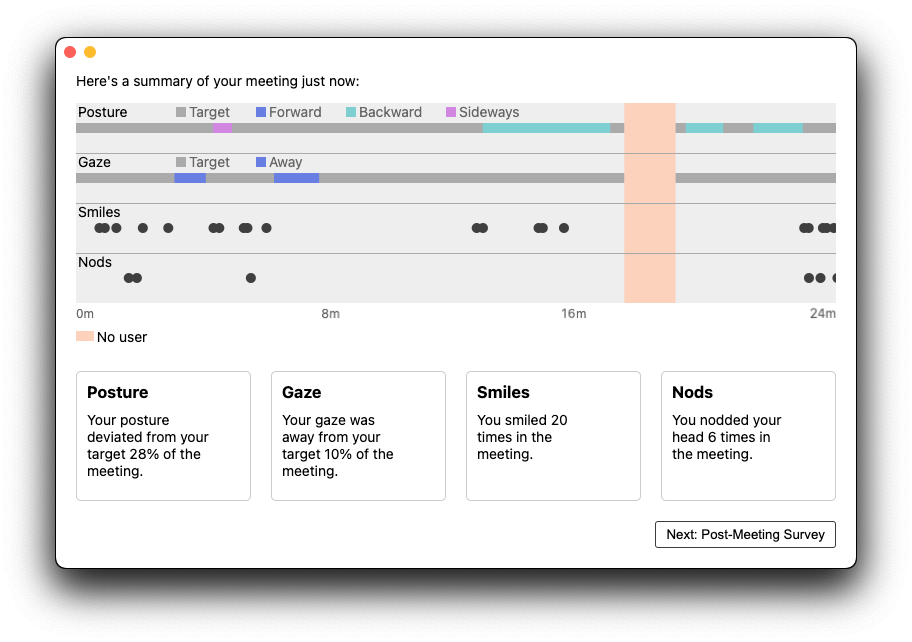}
  \vspace*{-9mm}
  \caption{The Post-Session Summary window, which contains a timeline visualization and summary statistics of the four non-verbal cues. This example indicates that the user had non-target gaze near the beginning, leaned backwards near the end, was out of frame for a short duration (shaded orange area), and smiled and nodded mostly at the beginning and end of the session.} %
  \label{fig:mt-visualization}
\end{figure}

\subsection{Post-Session Summary}\label{sec:postmeetingsummary}

When a \textit{Novecs} session ends, the Post-Session Summary window (see Figure~\ref{fig:mt-visualization}) displays a timeline visualization and summary statistics (proportion of non-target gaze and posture \& total smile and nod counts) for all four non-verbal cues. 
\rev{While the real-time feedback of non-verbal cues through the In-Session Sidebar is the primary focus of \revi{\textit{Novecs} as a technology probe}, \textit{Novecs} also includes \revi{intentionally minimal} summative feedback through the Post-Session Summary to explore how both types of feedback might be experienced by users in tangent.}

The visualization contains four horizontal timelines, one for each of the non-verbal cues, with the x-axis representing the spent meeting time (in minutes). 
Posture and gaze timelines are accompanied by a legend that defines the colour scheme; grey shaded parts of the bar represent target posture/gaze, while non-grey shading represents non-target posture/gaze.
The length of the shaded area corresponds to the duration of the detected non-verbal cue. 
Each detected smile and nod event is visualized as a black dot in their respective timelines.   
Shaded sections of the visualization that cut across all timelines represent time periods when \textit{Novecs} has been paused (light blue) or when the user is absent (light orange).

%% file: sections/4-method.tex
\section{Method}

\rev{We conducted an exploratory field study with the \textit{Novecs} technology probe to elicit knowledge workers' reflections on the system in-situ, and to investigate the potential opportunities and challenges of displaying real-time non-verbal cue feedback for supporting self-presentation during online work meetings.}
The study was conducted entirely remotely, but involved researchers and participants based in two countries on separate continents [countries omitted for anonymity]. %

\subsection{Participants}

Participants were recruited through public study recruitment platforms, social media posts, online study mailing lists, and convenience sampling through the research team's personal networks. To be eligible for the study, participants needed to: (a) be knowledge workers that attend at least 2-4 online meetings for work weekly with video on; (b) be using a computer running macOS 12.0 or higher for work, with at least 2 GB of free/unused RAM on average; and (c) use a webcam for video calls (including both built-in webcams, such as those on MacBooks, and external webcams). 

\rev{We recruited 30 participants who met our eligibility requirements through a screener survey. 12 participants dropped out, citing busyness, lack of qualifying meetings, or other unresolved technical issues. 18 participants (women = 9, men = 9, non-binary = 0; see supplementary materials for full demographics table) completed the study. They were on average 32 (SD: 8.83, min: 23, max: 51) years old at the time of the study. 9 participants were recruited from Switzerland and 9 from Canada. We used a broad definition of knowledge workers who came from a number of industries and backgrounds, including: Information Technology, Education, Administration, Engineering, and Medical Technology. 2/18 participants were PhD students working on their research (i.e., not taking courses), the remainder were professionals with roles including: Communications Lead, Account Manager, Marketing Associate, Data Analyst, and Product Designer. Participants were compensated with a 40 CHF (or local currency equivalent) gift card.}
\revi{The study was approved by the research ethics boards at the researchers' academic institutions in both countries.}

\begin{figure}[h]
\vspace{12pt}
  \centering
\includegraphics[width=.95\textwidth]{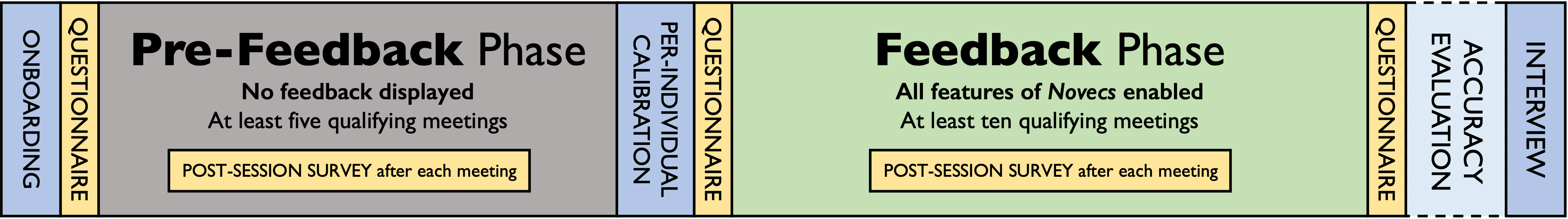}
  \caption{Diagram visualizing the study procedure.}
  \label{fig:mt-procedure}
\end{figure}

\subsection{Procedure} \label{sec:procedure}

In this section, we describe the study procedure for advancing participants through the two phases of our field study: \textbf{Pre-Feedback} and \textbf{Feedback} (see Figure~\ref{fig:mt-procedure} for an overview). 
We began by onboarding participants to the study through an initial 30-minute Zoom call, conducted by one of the lead researchers. In the onboarding session, participants were given an overview of the study procedure and goals, installed the Pre-Feedback phase version of \textit{Novecs}, and familiarized themselves with the tool, \revi{including how to start, pause, stop a meeting session, and guidelines for positioning the In-Session Sidebar appropriately, relative to video-conferencing window and their self-view.} Participants completed the pre-study questionnaire as soon as possible after the onboarding session and before starting the Pre-Feedback phase of the study.

During the \textbf{Pre-Feedback phase} of the study, participants continued attending video meetings as usual with \textit{Novecs} until they completed five or more qualifying sessions. Qualifying sessions needed to meet the following criteria: (a) the participant's video was enabled for more than 50\% of the meeting; (b) most meeting attendees were remote; (c) the participant self-reported that their overall level of interaction in the meeting was more than just passively listening; and (d) the meeting was at least 15 minutes long. These criteria were selected so that participants experienced \textit{Novecs} in meetings where it could potentially be relevant.

Whenever participants started a meeting session in \textit{Novecs}, a \textit{per-session} calibration step would be initiated (to determine ``target gaze and posture'' for that meeting, see Section~\ref{sec:inmeetingsidebar} for details). For the Pre-Feedback phase, \textit{Novecs} did not display any feedback to the participant, aside from detected user presence or absence. After finishing a meeting and ending the session, participants answered the post-session survey to report meeting context information, which was then used by \textit{Novecs} to determine if the past meeting was qualifying or not. 
The primary intention behind the Pre-Feedback phase was to give participants time to acclimate to manually starting \textit{Novecs} with their regular work meetings.

After the Pre-Feedback phase, participants completed the mid-study questionnaire and participated in another 30-minute Zoom call with a lead researcher to be onboarded to the \textbf{Feedback phase}, where both types of feedback, the In-Session Sidebar and Post-Session Summary, were enabled. In addition, participants were guided to complete the \textit{per-individual} calibration step, to tailor thresholds for each non-verbal cue specifically to the participant. The per-individual calibration only occurred at this point in the study to avoid influencing participants from focusing on the study's four non-verbal cues at the study outset; participants could not adjust these thresholds without researcher intervention; threshold changes were made only if significant issues arose with the non-verbal cue detection. 
New in the Feedback phase was that after completing the per-session calibration, participants would see the In-Session Sidebar. 
After finishing a meeting and ending the session, participants were first shown the Post-Session Summary before \revi{they were prompted to complete} the post-session survey.
Participants were part of the Feedback phase until completing ten or more qualifying (same criteria as in the Pre-Feedback phase) sessions, upon which they completed the post-study questionnaire and were invited to a final 60-minute Zoom call, consisting of a semi-structured interview, the accuracy evaluation step (for participants who opted-in, see Section~\ref{sec:accuracyEvaluation}), data transfer, debriefing and uninstallation of \textit{Novecs}. 

\subsection{Data Collection}
Throughout the study, we collected both qualitative and quantitative data via (a) the \textit{pre-}, \textit{mid-}, and \textit{post-}study questionnaires, (b) the post-session surveys, and (c) the semi-structured exit interview. 
Our primary focus was on the \textit{qualitative} data, due to the exploratory nature of a technology probe \cite{hutchinson2003technology} study; we include quantitative data analysis where appropriate to help contextualize our qualitative findings.
We conclude this section by describing the procedure for quantitatively assessing the sensing accuracy of \textit{Novecs}.

\subsubsection{Pre-, Mid-, and Post-Questionnaires}
The three questionnaires each contain the same set of questions (except demographic questions only asked in pre-study).
In the questionanires, participants rated their awareness of their own non-verbal cues during video meetings, the importance of non-verbal cues during video meetings for meeting effectiveness, and whether their own non-verbal cues matched those of an effective meeting participant. 
They also included Likert-scale questions to better understand participants' habits during video calls
as well as select questions of the Zoom Exhaustion \& Fatigue scale~\cite{fauville2021zoom} and Emotional Intelligence Scale (WLEIS)~\cite{wong2002wleis}; specific wording and analysis for this data can be found in the supplementary materials.

\subsubsection{Post-session Surveys}
After each session, the probe asked participants a few questions about the past meeting, such as: the number of meeting attendees, modality (hybrid or fully remote), video conferencing software used, whether screensharing was used, whether participants' shared their video feed, and their level of interaction (ranging from passive listening to active participation). Qualifying sessions included two additional 5-point Likert-scale questions on awareness of and satisfaction with their non-verbal cues during the past meeting. 

\subsubsection{Semi-structured Exit Interview}
We conducted 18 semi-structured interviews at the end of the study, starting by asking about their overall reactions to and general usage patterns with \textit{Novecs}, such as when and how often they noticed or glanced at the In-Session Sidebar. We also asked about potential learnings and impact from using \textit{Novecs}, to better understand the usefulness and usability of the In-Session Sidebar and Post-Session Summary. Participants were further asked to share their perspectives on each of the four non-verbal cues and use cases where \textit{Novecs} might be more or less valuable. We concluded by discussing \textit{Novecs'} design, including any suggested improvements or missing functionalities. Interviews were audio-recorded and transcribed for qualitative analysis.

\subsubsection{\textit{Novecs} Usage and Meeting Session Data} 
Whenever participants manually started a meeting session, \textit{Novecs} collected usage and meeting session data in the background. The collected data includes information on session length, \rev{raw data on when each of the four non-verbal cues was detected during a session}, and usage data, such as when the participant paused/resumed the session. No data specific to the meetings, such as meeting content, was collected.
Note that the quantitative data analysis is from 17 participants (P8's data aside from their interview was corrupted).

Overall, the quantitative usage data showed that participants varied in the duration of the time they participated in the study.
Participants took 1.43 weeks on average (SD: 1.41, max: 6.03, min: 0.29) in the Pre-Feedback phase (five qualifying meetings) and 3.71 weeks (SD: 2.93, max: 11.86, min: 0.86) in the Feedback phase (ten qualifying meetings). 
Participants used \textit{Novecs} for a total of 279 qualifying meetings, 106 in the Pre-Feedback phase (6.2 meetings on average per participant) and 173 in the Feedback phase (10.2 meetings on average per participant). 
Meeting session lengths were similar in both phases, with averages of 41.49 minutes (SD: 21.02, min: 15.98, max: 114.44) in the Pre-Feedback and 42.11 minutes (SD: 23.07, min: 15.15, max: 142.71) in the Feedback phase. 
176 meetings were held in Zoom, 51 in Microsoft Teams, 17 in Google Meet, 17 in DingTalk, 14 in Slack Huddles, and 4 in other video conferencing platforms. There were 130 meetings with 2 participants, 36 with 3, 50 with 4-5, 42 with 6-10, and 21 with 10+. Additional usage statistics can be found in the supplementary materials. 

\subsubsection{Accuracy Evaluation}\label{sec:accuracyEvaluation}
We quantitatively evaluated \textit{Novecs'} accuracy by asking participants to self-assess 64 images, sampled from the set of webcam images stored during meeting sessions.

\textit{Image Data Collection:}
To collect image data, \textit{Novecs} persisted images taken from participants' webcams during sessions only if they opted-in to the accuracy evaluation step, which all participants did (except one participant with a corrupted dataset). Normally, for privacy reasons, all images are immediately discarded after real-time processing. As it is too intrusive and resource intensive to store and label all image frames from a session, our approach was to store a specific, pre-defined subset of 60 image frames for each meeting and participant.
24 of these stored frames were based on random moments during the meeting. The remaining 36 frames (9 per feature) were selected based on 3 different ``moments of interest'', as calculated by \textit{Novecs}.
Since the goal was to assess accuracy, ``moments of interest'' were based on when maxima, minima, or challenging frames were detected, according to a strength score that was calculated by our detection approach and was specific to each of the four non-verbal cues. Using ``smiling'' as an example, 3 image frames were each selected from the session's pool of image frames, where the calculated ``smiling'' score was highest (maxima), lowest (minima), and closest to the threshold for differentiating between smile/non-smile (challenging).

\textit{Participant Assessment:}
As part of the exit interview, participants were instructed to rate a subset of their own collected image frames (of themselves; 64 in total) to produce labelled data that could be used as the quantitative accuracy evaluation.
We used a stratified random sampling method where 16 frames (4 random, 4 challenging, 4 min, 4 max) were randomly sampled each for all four non-verbal cues\revi{, across all meetings.}  
For each of the sampled image frames, we presented the frame (either as a still image or GIF for depicting the nodding motion), together with a positive or negative statement describing the state of the non-verbal cue.
For example, the statement would say ``You are smiling.'' for a maxima, ``You are not smiling.'' for a minima, and either statement randomly picked, for random and challenging frames. When reviewing these 64 image frames, participants could exclude images they were not comfortable sharing with the researchers. In total, we collected 1087 accuracy ratings (one participant's data only had 63 ratings).

\subsection{Data Analysis}
We analyzed interview transcripts using the Braun and Clarke approach to reflexive thematic analysis \cite{braunUsingthematicanalysis2006}, identifying themes with \revi{a primarily inductive approach, while being theoretically informed by existing literature in self-presentation and glanceable display design. Our codes and subsequent themes were constructed from the data and were not predetermined by an existing theoretical framework.} Three members of the research team first independently coded a subset of the interviews (3/18) to generate preliminary codes. Then, a lead researcher, paired with one of the two other team members, each coded separate subsets (3/18 each), meaning that half of our dataset (9/18) was coded by at least two members of the research team. One of the lead researchers then coded the remaining interviews (9/18).
Our data analysis process was marked by frequent meetings to review our codes and discuss a range of interpretations and perspectives in analyzing the data. In line with best practice in qualitative analysis, we did not calculate inter-rater reliability, deciding instead to use multiple coders to achieve `crystallization' \cite{ellingsonEngagingCrystallizationQualitative2009, tracyQualitativeQualityEight2010}. We took a recursive approach to our data collection and analysis process, moving back and forth between analyzing transcripts and iterating on the semi-structured exit interview questions. 
\revi{Although we reported participant counts per theme, we acknowledge (in line with Braun and Clarke \cite{braunUsingthematicanalysis2006}) the key limitations to this convention for representing prevalence. Counts mask the depth to which a participant might be articulating a theme and is often rendered problematic due to the nature of qualitative data collection \cite{braun2021thematic}.}
For quantitative analysis, we calculated and reported descriptive statistics from the Likert-scale questions in the pre-, mid-, and post-questionnaires and the post-session surveys to provide context to the qualitative findings, where appropriate. 

\textbf{Researcher positionality and reflexivity.}
We describe our positional stance as researchers to help contextualize the lenses by which we viewed the data and subsequent analysis. We reflected, discussed, and shared our own experiences with online meetings using various video conferencing platforms and our own challenges with expressing and interpreting non-verbal cues. In particular, three members of the research team work at an academic institution in Country1, and two in Country2; members from both countries were involved in coding, enabling us to share and \revi{contemplate} on our own cultural perspectives. \revii{Both of the lead researchers of the study} also experienced using \textit{Novecs} regularly in their own online work meetings. The research team consists of two graduate students, one senior researcher, and two professors. All members of the team are knowledge workers who regularly participate in online work meetings.

%% file: sections/5-findings.tex
\section{Findings}

To contextualize our findings, we first summarize participant perceptions of \textit{Novecs'} accuracy and calculated accuracy scores, suggesting that the technology probe was functional with participants' real meetings. We then delve into each of our four identified themes, beginning by highlighting participants' increased awareness of their non-verbal cues \textbf{(Theme 1, RQ3)}. 
Next, we describe the specific meeting types where non-verbal cue feedback was perceived to be most useful \textbf{(Theme 2, RQ1)}. We also identified tensions between authentic and in-authentic self-presentation, and how critical natural timing is when changing non-verbal cues \textbf{(Theme 3, RQ2 \& RQ3)}. While \textit{Novecs'} neutral feedback design helped with authenticity and natural timing, it also surfaced additional tensions in noticeability, distractibility, and actionability \textbf{(Theme 4, RQ3)}. 
We conclude with participants' overall willingness to continue using systems like \textit{Novecs} in the future. 

\subsection{Perceived Accuracy and Accuracy Scores}

\revv{When asked about accuracy, 7/15 (1 was not asked, and 2 were not confident in their awareness to provide a conclusive response) participants shared that they perceived \textit{Novecs} to be fairly accurate for all four non-verbal cues and 7/15 shared that it was accurate for some of the cues but not all.
}
1/15 (P3) felt like it was ``not quite accurate'', but also said that ``even though they have some false positives, I still think it is useful'' (P3), suggesting that they may have still been able to effectively experience the probe during the study. Inaccuracies around smiling were mentioned by 5/15 participants, nodding 4/15, gaze 3/15, and posture 2/15. Participants experienced occasional false positives with smiling and nodding, and gaze had issues when participants had wide screen monitor setups.
\revi{Unsurprisingly, when there were too many false positives, a few participants shared that they began to doubt or ignore some of the feedback for problematic cues, and selectively focused on cues that they believed to be more accurate, which would have shaped their experience of \textit{Novecs}. We took note of these cases, checked with participants during the interview to ensure that their reactions were not due to excessive inaccuracies, and also reviewed our selected quotes to provide additional context around accuracy if-needed.}
The quantitative accuracy evaluation of \textit{Novecs}' sensing accuracy at the end of the study (see section~\ref{sec:accuracyEvaluation}) revealed an average overall accuracy of 74\% across all four types of cues. %
The individual average accuracy scores for each non-verbal cue were 67\% for gaze, 72\% for nodding, 71\% for posture, and 83\% for smiling. While there are differences in detection algorithms along with how and where (i.e., in-situ versus lab) accuracy is evaluated, these scores generally fall in line with existing literature on automatic non-verbal cue detection, which mostly hover between 65-85\% \cite{girardiEmotionsPerceivedProductivity2021,baltrusaitis2018openface,alghowinem2013eye}.

\subsection{Theme 1: Increased Awareness of Non-Verbal Cues during Video Meetings}
\label{sec:themeIncreasedAwareness}

Overall, 13/18 participants shared that \textit{Novecs} helped increase their awareness of their non-verbal cues. Participants attributed these awareness increases to both the real-time feedback, which helped with \textit{in-the-moment} awareness, and also the post-session feedback, which helped with \textit{recapping} the meeting. While the impact was mainly on awareness, a few participants also reported short-term behaviour changes in their non-verbal cues during meetings.

\subsubsection{Real-time Feedback Facilitated In-the-Moment Awareness}

\rev{We begin by summarizing participants' different approaches to becoming aware of the real-time feedback in meetings. 10/18 participants self-reported a \textit{proactive} approach of glancing at the sidebar regularly to check on the feedback:
``Maybe like every five to ten minutes or so, I would probably just take a look or maybe if I wasn't actively talking for a while, I would probably take a look at it'' (P13). Others (11/18) \textit{reactively} noticed in their peripheral vision when the sidebar's colours changed as a signal, without actively glancing at it: ``[The sidebar] would just kind of catch my eye every time one of those widgets (...) turned grey, but otherwise, yeah I was kind of looking at the counter for like the last time I smiled or nodded. Basically, every time the widget changed, I'd kind of glance over at it'' (P2).}

9/18 participants in the study 
reported how the real-time feedback helped them to think about their own non-verbal cues ``in-the-moment'' (P17). Specifically, the \textit{Novecs} sidebar allowed them to become aware of, and reflect on, the \textbf{moment-specific appropriateness of their non-verbal cues during the meeting.} Rather than simply being aware of their non-verbal cues in general, participants were able to connect their behaviours to that moment in the meeting, and reflect on if it was appropriate or not. P10 used the real-time feedback to make sure that they were ``showing the right reaction to something'' and if their expressions were ``matching what's going on''. P10 added, describing a specific example where they changed the amount of times they smiled to be less to match that moment in the meeting: ``I felt like I reduced that a bit, and like where it's more appropriate to do that, just kind of thinking about matching more towards the kind of overall atmosphere of the meeting.'' Similarly, when P9 was asked about how they reacted when they noticed the real-time feedback about smiling, they said: ``Like my [role] was sort of a shadow figure in the meeting, it was actively listening, but not actively participating, so probably it makes a bit of sense [to not smile]. The tool helped me realize that, I think I would not have noticed that by myself probably.''

\subsubsection{Post-Session Feedback Helped Participants Recap and Reflect}
\label{sec:postMeetingAwareness}

10/18 participants 
liked the post-session feedback as a way to recap the meeting that had just occurred. 
Specifically, the post-session feedback helped participants become \textbf{more} \textbf{aware of and reflect on their participation and engagement} in the meeting by looking at the non-verbal cues that were visualized on the timeline. P2, for example, used the empty spaces in the timeline to become aware of when they might have been less engaged: ``It showed me when I was kind of zoning out in a meeting or something when there were some sections that were like, completely bare.''
Similarly, P10 reported the summary helping to reflect on the meeting outcome: ``I thought it was pretty cool to like see over time (...) and also think of how the discussion had unfolded and how that matched up with the different kinds of movements like nodding and stuff like that.''

Participants further stated that the post-session feedback allowed them to \textbf{better understand their behaviour depending on the meeting and sort-out misconceptions} they had about their own expressions. For example, P9 became more aware of how their nodding and smiling changed in a programming session compared to other meetings: ``I was nodding way more often, I was nodding a lot because he was explaining to me a lot of things... And it was like a bit less smiling, it was a technical session so it was less smiling. While at other meetings that are a bit more relaxed, I was smiling more and nodding less. So it was mostly for me a confirmation of the type of meeting I'm in and what's my reactions to it.'' Similarly, P15 realized that they were more expressive during meetings than they had previously considered: ``When I have to rate myself, I would always be more on the negative side. If you asked me how my facial expressions are during meetings before this study, I would have said it's probably really bad because communication is not my strength. But then seeing that I actually nodded and smiled quite a lot made me feel that maybe it's not as bad as I thought.''

\subsubsection{Impact on Short-term Behaviour Change}

8/18 participants 
reported short-term behaviour change during meetings, such as how the real-time feedback helped to ``adjust and correct behaviours'' (P18) and ``course correct in the middle of [meetings]'' (P17).
P6 also hinted at longer-term behaviour change, saying: ``But like for sure the leaning back thing, I keep coming back to that, it's like ingrained in me now. Like honestly I do it myself or if I go too far back, I see it now on the screen and it comes forward.'' However, 6/18 participants (P1, P2, P7, P9, P12, P17) also shared that it didn't ``change much [of] my behaviour during the meetings'' (P9). \revi{This was consistent with the quantitative data from the post-session surveys and study questionnaires that measured whether or not participants' assessment of their non-verbal self-presentation changed after the study.}
Our t-test ($\alpha=.05$) showed no significant differences in participants' self-reported satisfaction (5-point Likert scale, 5 denotes extremely satisfied) with their non-verbal cues when comparing the Pre-Feedback (mean: 3.46, SD: 0.74) and Feedback phase (mean: 3.43, SD: 0.75) from their post-session surveys. Similarly, no differences were found from participants' ratings in the study questionnaires (5-point Likert scale, 5 denotes increased agreement) on whether their non-verbal cues matched that of an effective meeting participant (means for pre: 3.72, mid: 3.94, post: 3.94). 
See supplementary materials for further details of this analysis. 
While we did not expect to see \revi{measurable, long-term behaviour change in this study, as it is ultimately a long-term process \cite{klasnjaHowevaluatetechnologies2011}, that almost half of the participants self-reported \revii{some} behaviour change} \revii{hints} at the potential of a real-time, non-verbal cue feedback system.

\subsection{Theme 2: Different Meeting Types Impact the Utility of Non-verbal Cue Feedback}
\label{sec:themeUsefulTypes}

\rev{We identified how different types of meetings can influence participants' perceived utility of real-time non-verbal cue feedback. Meeting types differed in factors such as meeting size, the importance of the meeting, and familiarity with the other meeting attendees.}

\subsubsection{Number of Meeting Attendees}

The number of meeting attendees influences how active and engaged participants are or feel they need to be, and thus, \textbf{impacts their cognitive capacity to pay attention to, and act on, the feedback} provided through the sidebar. 12/18 participants reported that the real-time feedback was generally most useful in small-to-medium sized group meetings when they were not in a constant state of active participation, but still need to be engaged:
``In meetings, where there's more people and my role is not that active, I think it's useful. Especially there are some meetings when I'm really, really trying to take as many notes as possible and I don't say much. And then I concentrate a lot on writing. And then from time to time seeing `ohh, I should actually show that I'm still here.''' (P15).

To the contrary, feedback in one-on-one meetings was generally considered as less appropriate, since participants were talking for most of the time and had little-to-no cognitive capacity to pay attention to the sidebar: ``Generally for like one-on-one meetings where like I'm talking 50\% of the time, then I'm not really paying attention to the widget... maybe not so much when I'm talking, but when other people are talking. Obviously I have more capacity to take a look at the tool then.'' (P2).
Some participants (5/18: P2, P5, P10, P13, P16) also shared that the rapid, back-and-forth conversational pace in one-on-one meetings sufficiently signaled engagement.  %

Unsurprisingly, for meetings with a large number of attendees, the feedback from \textit{Novecs} was less useful as their non-verbal self-presentation didn't matter: ``If it's a session with 70 participants and I'm one of many small pictures on the screen... then I don't care how I come across'' (P7).

\subsubsection{Importance and Formality of the Meeting}

7/18 participants described that the non-verbal cue feedback was particularly useful in \textbf{meetings that had more at stake in terms of importance or were more formal}. \revi{Participants described how examples of such meetings were ones with external clients or supervisors, where there were greater potential job repercussions, compared to, say, a meeting with their regular team. Participants felt like such meetings were ones where they had to be especially aware of and better manage their non-verbal cues.} 
For example, P6 said: ``It's like what is the level of professionalism needed? That's kind of what I used it [\textit{Novecs}] for -- to help me.'' 
P3 discussed how \revi{their perception of how important the other attendees in the meeting affected how conscious they were about their non-verbal cues}: ``When I have the meeting with important roles then I have to adjust my visual language and body language based on the tool [Novecs].''
In addition, when asking participants about how they would see themselves using \textit{Novecs} in the future, a few participants (3/18: P7, P8, P17) identified that they would find the tool particularly useful when training or onboarding people as well as when interviewing for jobs, which emphasizes the utility of \textit{Novecs} in more \revi{consequential settings, where impressions might matter more}. 

\subsubsection{Familiarity with Other Meeting Attendees}

Related to formality is participants' familiarity with the other meeting attendees. 8/18 participants felt that it was \textbf{less important to manage their non-verbal cues with meeting attendees they were already familiar with}. For example, P4 said: ``The people I'm working with and my team, I already know them really, really well, so I don't feel the need to be super polite and nice and whatever. So I think that I usually nod more or smile more if it's a new person.'' Although they were aware of ``the whole idea of being forward to show interest'', P18 was comfortable with leaning back only because they were with ``people that I really felt super comfortable with, and [were] less focused on (...) the presentation of myself.'' In these meetings, familiarity allowed participants to feel like they could be themselves and not have to manage their self-presentation, which is discussed further in the next section.

\subsection{Theme 3: Tensions between Authenticity, In-Authenticity, and Natural Timing in Self-Presentation}\label{sec:tensionAuthenticitySelfPresentation}

Participants emphasized the importance of being authentic in how they expressed themselves in work meetings, but recognized that there are some situations where they had to externally present themselves in a manner different from how they were feeling internally (i.e., in-authentic self-presentation \cite{peck2018acting,bolino2016impression}). %
They further wanted their self-presentation efforts to appear natural and with appropriate timing.
\revv{\textit{Novecs'} neutrally-framed feedback was generally viewed as compatible with navigating this tension and the need for natural timing. 
}

\subsubsection{Importance of Authentic Self-Presentation}

When it comes to self-presentation in work meetings, participants \textbf{stressed the importance of being ``authentic''} (P8), ``genuine'' (P2), ``natural'' (P2, P3, P11, P17)\revii{, and ``organic'' (P6).}
A few participants (4/18: P2, P3, P4, P17) even alluded to the importance of embracing ``humanness'' (P2) and imperfections. P17, for example, emphasized how being authentic sometimes means displaying imperfect non-verbal cues: ``When you have a lot of meetings like me in a day, I mean, I'm not like a robot. I am going to move. I am going to look around. I am going to get bored. I am gonna (...) sometimes not be in the mood (...). And so I don't believe there's a perfect meeting behaviour. Everybody has their own personalities, their own characteristics, and I think my philosophy is like you can make faces, you don't have to have the perfect [expression] like you could have a resting whatever face, but as long as you're answering the questions and being respectful and you're actively listening, engaging, et cetera [sic], then (...) you can overlook some of the human things like that.''

\revv{P8 explained how the In-Session Sidebar of \textit{Novecs} was compatible with their emphasis on authentic non-verbal behaviour, both of themselves and of others: ``There's still a way to balance being your authentic self while also using the sidebar for cues... I just want them \revii{[others]} to be how they are. They \revii{[others]} don’t have to go out of their way to do all these extra things very frequently. Especially, if, that's just not who they are.''
They continued, saying that the sidebar was a reminder of their current behaviour, but never forced them to change from their authentic self: ``I wouldn't correct it unless it felt right to do it. And that's why I just kept my body language and my gaze the same because that was authentic to me (...) It didn't feel like I had to go out of my way to smile extra or not [smile] extra. It was just a little reminder of `ohh, you haven't [smiled] in a while''' (P8).
}

\subsubsection{In-Authentic Self-Presentation is Occasionally Necessary}

Despite the importance of authenticity, a few participants (5/18: P3, P8, P12, P13, P14) also explicitly mentioned how they needed to engage in in-authentic self-presentation, where they had to \textbf{outwardly express something different from what they were feeling internally}. 
Different meeting types may help explain situations where authentic or in-authentic self-presentation is needed (e.g., important or formal meetings may require more in-authentic self-presentation; see section~\ref{sec:themeUsefulTypes}).

\revv{In situations that call for in-authentic self-presentation, participants like P3 and P8 explicitly describe how \textit{Novecs} was a useful reminder. P3 mentioned a few meetings where they needed to smile more to express positivity and agreement more than they felt internally, finding \textit{Novecs} to be ``very useful'', saying: ``[When] my supervisor asks me to do something even though I'm not happy, but [sic] I pretend to be happy. I pretend, to hide my actual emotion (...) when I have a meeting with the master's student who helped me perform experiments, even though I'm not so satisfied, I should still smile a lot.''}

P8 also shared a story of when they were having a bad day but still had to facilitate a session at work, another example of an important meeting \revv{where in-authenticity might be necessary}: ``When you are facilitating you can't really be your authentic self, you have to put on that little customer service vibe... I wasn't feeling it, but I also couldn't completely be my authentic self... It was like a good reminder --- I haven't done that in a while, and I'm talking to 30 people, so I should smile a little, or look a little more approachable, which is something I found it [\textit{Novecs}] really helpful for actually, because I struggle with looking approachable, and looking friendly, and showing some of those indicators in larger groups.'' While not explicitly discussed by participants in this manner, we note that in-authentic self-presentation may also occur due to a lack of self-awareness of their non-verbal cues (i.e., unconsciously \cite{bolino2016impression}). For example, if one feels agreeable internally, but forgets to express it externally through nodding or smiling; our study design could not capture this. 

\subsubsection{Natural and Appropriate Timing is Critical}

A few participants (4/18: P6, P8, P11, P13) stressed the importance of \textbf{natural and appropriate timing for changing their non-verbal cues}, as opposed to simply doing so whenever they noticed feedback signals. 
\revv{
For example, P6 mentioned: ``Nodding is like, if someone makes a point, you want to give them an affirmation of that point. But I wanted to make sure that it was organically done and not me just being like I haven't nodded in two minutes... I do try and do that and this study did bring that to light.'' 
P13 described their thought process when seeing the feedback as waiting for the next opportune moment, saying: ``if it says I didn't smile for [some time], I would try to smile at the next opportunity.''
}

6/18 participants (P2, P5, P8, P11, P13, P16) also called out the intentionally neutral, \rev{non-normative} feedback that \textit{Novecs provided as compatible} with supporting natural timing for changing their non-verbal cues in their work meetings. 
\revv{
For example, P11 observed how the ``prompt doesn't ask you to smile, it just says that you haven't smiled for these many minutes'', saying that they should smile according to the ``natural flow [of conversation]''. 
Similarly, P6 shared how the In-Session Sidebar's last smiled/nodded minute counter did not pressure them to change their non-verbal behaviour ``without it feeling like the appropriate time'' despite the ``timer getting high.''
}
\rev{However, P2 and P17 felt like the reminder could sometimes \revv{influence them} to change in unnatural ways.
Although P2 said that the tool was ``not so much pressure'' and was ``not like I was being forced to do that'', they did describe how it felt ``a little ingenuine at times''. When referring to the last smiled minute counter, P2 described: ``I was trying to keep that number down, so I would kind of smile or nod more frequently, even if it didn't feel natural to the conversation that was being had.''}

\subsection{Theme 4: Tensions between Noticeability, Distractibility and Actionability}

While \rev{non-normative} and neutral feedback was one of the key design goals of \textit{Novecs}, and could be important for supporting natural timing of changing non-verbal cues (as suggested by Theme 3), it appears to have also surfaced tensions around balancing noticeability, distractibility and actionability for real-time displays of non-verbal cues. 

6/18 participants (P4, P8, P10, P12, P15, P16) explicitly noted that the real-time feedback was ``subtle'' (P8, P16) and ``minimal'' (P12), and thus did \textbf{not distract them from focusing on their meetings}. Noticeability and distractibility are a known tradeoff, as increasing the saliency of the feedback would also draw users' attention away from their meetings. 
The neutral colour scheme of visualizing data in a light blue or grey shade, as opposed to e.g., a bright red shade, likely added to the non-distracting experience.  %
For example, P4 shared: ``I never felt interrupted from the tool during a meeting... I was not forced to look at the tool and I really like that because I feel if I had like a red light popping up or something, I would always be distracted because I'm easily distracted during meetings sometimes. So I really liked that it's subtle.'' Similarly, P16 said: ``It helped that the colour is not (...) that interruptive. So it's not like a red signal blinking when I'm leaning a little bit behind. (...) I've never felt judged by the tool.'' \rev{Beyond not being distracting, we also note that P16 brings up not ``feeling judged'' by the colour scheme, which is consistent with our goal of avoiding overly normative feedback design.}
\rev{\revv{When viewing the In-Session Sidebar,} 6/18 participants (P2, P6, P9, P14, P15, P18) also suggested that the real-time feedback by \textit{Novecs} could be \textbf{more actionable, such as by being more explicit about what actions to take}. \revv{Interestingly, some participants suggested adding ``high-level'' constructs like engagement, attention, or emotions, despite known challenges with actionability for such constructs \cite{hartzler2014real}.} P9 wanted a ``facial emotion detector (...) and to see if (...) half of the time you have a pissed off expression, or a happy expression.'' It may be that ``higher-level constructs'' help provide a rationale for changing one's non-verbal behaviour, whereas simply seeing, e.g., ``non-target posture'', may not spur users to action.   
In addition, P14 suggested wanting the tool tell them if their current posture was ``good or bad'' and ``how it might be interpreted by people seeing me'', suggesting some individual desire for more normative feedback.}

5/18 participants (P4, P5, P14, P15, P17) also expressed challenges with actionability when viewing the Post-Session Summary. While it was valuable to increase their awareness and reflect on their participation and engagement (see section~\ref{sec:postMeetingAwareness}), many did not know how to interpret and act on the presented feedback. More specifically, P4 shared that they struggled with \textbf{not having a point of reference} of how often one should be expressing non-verbal cues in a meeting: ``I don't really know what the bar should be like. How often should a person be smiling, or how often should a person be nodding in order to look like an engaged participant in a meeting?'' (P4). 
Other participants (P2, P8, P9, P18) also suggested having a way to compare their facial expressions and body language across different types of meetings (Theme 2, see section~\ref{sec:themeUsefulTypes}), which may help alleviate the lack of a reference point. For example, P2 stated: ``It would be helpful to tie the data into what kind of meeting I was in because my one-on-ones are a lot more engaged compared to a meeting that has seven people. So that data is kind of hard to compare. So if I see like side-by-side like `[at] my all-hands meeting (...) you were leaning back most of the time'. Whereas like in this [other] meeting you're engaged the whole time.''

\subsection{Willingness for Future Use}

Finally, the majority of participants (15/17, one did not provide a conclusive response) answered that they would consider using a system akin to \textit{Novecs} in the future. Of these, 10/17 were interested in continuing using \textit{Novecs} regularly as-is, 3/17 occasionally, 2/17 for a short period of time (e.g., for awareness training), and only 2/17 (P2, P7) did not want to use it further. P6 was particularly enthusiastic: ``I did really enjoy using it. I think it's very helpful for being engaged. It would be really good if it was built into some of the video conferencing software like that would be the ideal.''
\revv{Reasons for not wanting to continue using \textit{Novecs} were that they were already satisfied with their existing awareness (P7) and that being consciously reminded of their own non-verbal cues might lead to ``more burnout'' (P2).}

%% file: sections/6-discussion.tex
\section{Discussion}

\revv{
We begin by summarizing the key findings from the \textit{Novecs} technology probe. Next, we reflect on the importance of authenticity and the imperfect ``humanness'' in non-verbal self-presentation as well as the tradeoff between noticeability and distractibility, before concluding with design opportunities for future real-time, non-verbal cue feedback systems and study limitations.
}

\subsection{Summary of Key Findings}

\revv{
We learned that real-time feedback systems, like \textit{Novecs}, have the potential to foster self-awareness of non-verbal cues. Specifically, it enabled users to connect their behaviour with that current moment in the meeting, encouraging reflection on the in-the-moment appropriateness. When changing their non-verbal cues, participants stressed that this needed to be done with natural timing.
The neutrally-framed, real-time feedback encouraged reflection rather than immediate action in-the-moment, surfacing both tensions of authentic and in-authentic self-presentation as well as distractibility, noticeability, and actionability. Notably, we identified meeting \rev{types} where real-time feedback on non-verbal cues was most useful, such as small-to-medium sized meetings that involve unfamiliar meeting attendees and meetings that are deemed more important. 
}

\subsection{Importance of Authenticity and Imperfect ``Humanness''}

\rev{For online work meetings, participants shared that in general, they valued authentic non-verbal self-presentation, but acknowledged that there were also times where in-authentic self-presentation \cite{peck2018acting,bolino2016impression} was necessary, such as when there was a need to maintain a level of professionalism or when taking on a particular role in a meeting. 
Our findings extend the known tensions around authenticity and in-authenticity as identified by Goffman \cite{goffman2002presentation} through characterizing this tension specifically for knowledge workers in online work meetings. In addition, our findings on meeting types can potentially explain the specific scenarios in which authenticity is valued (e.g., when you are familiar with the other meeting attendees) or when in-authenticity is needed (e.g., in important or more formal meeting contexts).
\textit{Novecs} was compatible with navigating this tension, as well as supporting participants in changing their non-verbal cues with natural timing. 
\revii{
While reflection on in-the-moment appropriateness still requires intentional effort even with \textit{Novecs} support, long-term training could shift self-presentation from conscious to automatic \cite{peck2018acting}, eventually opening up the possibility of no longer needing to rely on non-verbal cue feedback.}
}

Notably, there may also be value in imperfect non-verbal cues in online meetings~\cite{isaacs1993video, brubaker2012, sauppe2014}, especially as some participants stated that they were comfortable with expressing their natural, imperfect self with familiar coworkers. 
It may also be reasonable to assume that imperfect non-verbal cues could build connection through expressing vulnerability or signaling one's well-being. For example, if one is feeling unwell or unhappy, authentically reflecting it in one's non-verbal cues through a lack of smiling or a slouched posture could be a signal to other meeting attendees to ask how they are doing, given a culture of trust and psychological safety \cite{edmondsonPsychologicalSafetyLearning1999}.

\subsection{Balancing Noticeability and Distractibility}

Our findings also replicate the known tensions around distractibility and noticeability from the glanceable display literature \revi{\cite{morrison-smithAmbiTeamProvidingTeam2021,tanveer2015rhema,terken2010multimodal,balaam2011enhancing}} but in the context of online work meetings.
\revi{In online work meetings, non-verbal cues often only take a secondary role to the actual work being conducted or discussed, unlike contexts such as public speaking \cite{tanveer2015rhema} or social conversations \cite{aliSocialskillstraining2017, aliLISSALiveInteractive2015}, where appropriate non-verbal cue behaviour may be more central.}
\revi{Another challenge in work meetings is managing the dynamics within the meeting. For example, workers may place greater importance on non-verbal cues more at the beginning or end of a meeting when exchanging pleasantries, whereas these cues may be less valued when meeting attendees are all focused on reading a screen-shared document. Distractability greatly fluctuates even within a particular work meeting.}
\revi{In our findings, w}hile \textit{Novecs} positively impacted participants' awareness on non-verbal cues and, in a few cases, even impacted short-term behaviour, several participants stated that they did not notice the glanceable display \rev{during some types of meetings.} While it was an explicit, and appreciated, design decision to avoid making the display too intrusive to distract from the actual meeting content, the impact that such a real-time display can have on awareness and behaviour is limited by it being noticed at the specific time of the feedback.

\rev{The type of meeting is one such factor to consider in balancing this tension. For example, in one-on-one meetings where participants often failed to notice the sidebar due to rapid back-and-forth discussions, one might consider only disrupting the user to notify them if there is a major issue with their non-verbal self-presentation (what constitutes a major issue may be defined by the user themselves), such as if they were leaning to the side for a significant amount of time in the meeting.}
\revi{Future work could also further consider the role of the user in the meeting (e.g., presenter, speaker, listener) in this specific tension and in the broader utility of a real-time non-verbal cue feedback system.}
Another way to increase noticeability while keeping distractions low could be to adapt the saliency (i.e., noticeability) of the feedback based on cognitive load. For example, feedback could be displayed only when the user is not speaking and can actually notice and act on the feedback. Designers could further customize the modality, such as feedback in the form of just-in-time notifications inside the video conferencing platform, rather than through a persistent, glanceable display (P2, also in \cite{tanveer2015rhema}), or even using audio (P4, P6) and haptic (P14) channels.

\subsection{Design Opportunities for Future Non-verbal Cue Feedback Systems}

\revv{
We propose the following design opportunities for future similar real-time, non-verbal cue systems for online work meetings. This list provides a starting point for future designers' consideration, in line with a technology probe's ``design goal of inspiring users and designers to think of new kinds of technology to support their needs and desires'' \cite{hutchinson2003technology}. We note that they need to be further iterated on and validated in more rigorous, larger-scale, and longitudinal studies. 
\begin{itemize}
  \item \textbf{Adapt the feedback design of such systems to meeting types:} Our findings suggest that small-to-medium sized meetings of importance with unfamiliar meeting attendees are meeting types where real-time non-verbal cue feedback are most useful. The saliency, noticeability, and distractability of feedback could adapt to the meeting type. \revi{For example, feedback on non-verbal cues could \revii{be more proactive or aggressive in notifying users in important meetings where maintaining their ideal non-verbal self-presentation is especially relevant.} On the other hand, meetings with a large number of attendees or with close team members could be ones where non-verbal cue feedback has reduced saliency or even larger thresholds for non-target behaviour.}
  \item \textbf{Encourage reflection on authenticity and in-authenticity:} Meeting types could also help characterize scenarios where authentic or in-authentic self-presentation might be needed. Systems could help users reflect on this dimension of authenticity in non-verbal self-presentation. For example, if a worker is leading a presentation for a meeting, they might realize that they need to take on a (potentially in-authentic) role of a confident and assertive presenter; Persona-like presets in future systems could be one way to encourage reflection and help target specific non-verbal cue behaviour for training.
  \item \textbf{Leave room for imperfection and avoid overly normative design by default:} Interestingly, a few participants emphasized not wanting to have to always manage their non-verbal cues well, and that it's okay to be imperfect, as it feels more ``human'', especially among close colleagues. Future designers should, like in \textit{Novecs}, avoid overly normative design; they could consider features that snooze or dismiss non-verbal cue feedback, \revi{or even prevent over-reliance by encouraging tapering use of the system once users are satisfied with their non-verbal self-presentation.}
  Conversely, a few participants desired more normative feedback to help with actionability; customization through a ``normativity slider'' could be a possible design solution.  
  \item \textbf{While low-level cues are more actionable than ``high-level'' constructs, some users want more guidance:} Participants struggled with not having a point of reference for expressing non-verbal cues. As non-verbal self-presentation is personal, systems like \textit{Novecs} can help knowledge workers begin to become more aware of their non-verbal cues and reflect on it, so that they can come up with personal improvement goals (e.g., smiling more often) that can serve as a personal point of reference to facilitate actionability. 
  Curated personas, as mentioned earlier, that can adapt to both the individual and meeting type, could also help with actionability, by serving as a point of reference for users to emulate in their non-verbal self-presentation. Personas may also be a way to reconcile constructs with cues; individualized targets for non-verbal cues could be tied with personas that are associated with constructs (e.g., the Active Audience persona demonstrates engagement by having target posture for at least 150\% more than your typical target posture time, but only for presentation meetings). Future work could also explore the evaluation of one's own non-verbal behaviour from the perspective of the other attendees to inform actionability. 
\end{itemize}
}

\subsection{Limitations}

\textbf{Privacy Considerations \& Self-Selection:} Due to the potentially intrusive nature of the approach, privacy was a key consideration. To ameliorate concerns, data processing for \textit{Novecs} was done locally and sessions were started manually. 
While we did not explicitly prompt participants to think about potential privacy concerns, no participants mentioned any concerns around privacy or feeling uncomfortable when using it. %
Although one reason may partially be due to self-selection biases \cite{heckman1990selection}, it is encouraging that people seem to be open towards using systems that leverage potentially sensitive facial data so long as they are designed to be privacy-preserving.

\textbf{Accuracy and Cultural Considerations:} A key limitation of any vision-based detection system is accuracy and the technologies it relies on~\cite{baltrusaitis2018openface}. 
\revi{Challenges with accuracy are likely exacerbated in deploying such a system in the field, with a variety of devices and environments where online work meetings would take place.}
\revi{A key issue is the experience of individuals with darker skin tones when it comes to such systems \cite{buolamwini2018gender}.} Unfortunately, while the study included individuals of various racial backgrounds, including Caucasian, East Asian, South Asian descent, we did not have any participants who identified as Black or Hispanic. While our study sampled participants from two countries with distinct cultures --- [countries omitted for anonymity] --- we did not observe noticeable differences in our findings between participants from either location. %
Cultural considerations were not a key factor in our study, but should be further explored, as there are known cultural differences in how people express themselves by their non-verbal cues \cite{bitti2011nonverbal, matsumoto2016cultural}. %

\textbf{Generalizability:} \revii{The sampling and number of participants, our choice of non-verbal cues,} and especially the types and frequency of the meetings with \textit{Novecs} enabled, might have impacted the generalizability of our findings. %
Findings on awareness impact may be influenced by the Hawthorne effect \cite{MCCAMBRIDGE2014267}. However, the aim of this exploratory work was not to achieve generalizability, but instead learn about participants' openness towards, as well as opportunities and challenges for real-time feedback. 
\revii{Further, the \textit{Novecs} technology probe only supported four specific non-verbal cues. However, they are not necessarily the only cues that future \textit{Novecs}-like systems should adopt, nor are they necessarily the ones that knowledge workers will find the most relevant in the work meeting context. The four cues simply served as a reasonable starting point for the \textit{Novecs} technology probe. \reva{We also acknowledge that by including smiling and nodding (but not frowning or head shakes), we focus \textit{Novecs} on capturing positivity and engagement. A future version could also consider negative non-verbal cues to capture confusion, frustration, or disagreement, but additional consideration would be needed in its design.}}

%% file: sections/7-conclusion.tex
\section{Conclusion}

Our work improves the understanding of how real-time feedback systems might support non-verbal self-presentation, as well as the potential opportunities and challenges of such systems for online meetings.
To that purpose, we deployed \textit{Novecs} in an exploratory field study with 18 knowledge workers.
\textit{Novecs} is a technology probe of a feedback system that automatically detects and visualizes users' non-verbal cues \rev{during online work meetings via both real-time and summative approaches.}
\rev{We learned that participants' awareness of their non-verbal cues generally increased and that there are tensions around authenticity and in-authenticity, which \textit{Novecs}' neutral feedback design was compatible with navigating.} We also identified meeting \rev{types} where real-time feedback on non-verbal cues was most useful.
\rev{In an era of hybrid work, where online meetings are ubiquitous, the findings and design opportunities surfaced by \textit{Novecs} can both inform and inspire future designers of such systems to empower users in enhancing their non-verbal self-presentation.}